\documentclass[showpacs,preprintnumbers,amsmath,amssymb,twocolumn,prb]{revtex4}
\usepackage{mathrsfs}
\usepackage{array}
\usepackage{amsmath}
\usepackage{graphicx}
\usepackage{amstext}
\usepackage{amsfonts}
\usepackage{bm}

\begin{document}
\title{Transient Current-Current Correlations and Noise Spectra}
\author{Pei-Yun Yang}
\affiliation{Department of Physics and Center for Quantum
information Science, National Cheng Kung University, Tainan 70101,
Taiwan}
\author{Chuan-Yu Lin}
\affiliation{Department of Physics and Center for Quantum
information Science, National Cheng Kung University, Tainan 70101,
Taiwan}
\author{Wei-Min Zhang}
\email{wzhang@mail.ncku.edu.tw} \affiliation{Department of Physics
and Center for Quantum information Science, National Cheng Kung
University, Tainan 70101, Taiwan}
\begin{abstract}
In this paper, we present an exact formalism for transient current-current correlations
and transient noise spectra. The exact solution of  transient current correlations
in both the time domain and the frequency domain are obtained.  
Without taking the wide band limit, we investigate transient current-current 
correlations with different bias voltages and different finite temperatures.
Transient noise spectra over the whole frequency range are
calculated and that in the steady-state limit are also reproduced.  From transient
current-current correlations and noise spectra, we analyze the frequency-dependence
of electron transport for the system evolving far away from
equilibrium to the steady state.  Various time scales associated with the energy
structure of the nanosystem are also obtained from the transient current-current
correlations and transient noise spectra.
\end{abstract}

\pacs{73.63.-b; 72.70.+m; 72.10.Bg; 85.35.-p} \maketitle

\section{Introduction}
Noise spectra provide the information of  temporal correlations
between individual electron transport events. It has been shown that
noise spectra can be a powerful tool to reveal different possible
mechanisms which are not accessible by the mean current measurement.
Examples include the information on electron
kinetics,\cite{Landauer39265898} quantum statistics of charge
carriers,\cite{Beenakker563703} correlations of electronic wave
functions,\cite{Buttiker81276398} and effective quasiparticles
charges.\cite{Saminadayar79252697, Quirion9006700203} Noise spectra
can also be used to reconstruct quantum states via a series of
measurements known as quantum state tomography.\cite{
Buttiker7304130506} Conventionally, evaluations of noise spectra are
largely limited to the rather low frequency ($\hbar\omega\ll k_BT$),
where the noise spectrum is symmetric at zero bias.\cite{Scatter2}
However, experimental measurements of high frequency noise
spectra\cite{Schoelkopf783370,Deblock30120303,Kouwenhoven9617660106,Zakka9923680307}
inspired the exploration of the frequency-resolved noise spectrum both
in symmetric\cite{Aguado9220660104,Lambert7504534007,Wu8107530910}
and asymmetric
form.\cite{Engel9313660204,Wohlman7519330807,Wohlman7907530709,Orth8612532412}
The asymmetric noise spectrum, which is directly proportional to the emission-absorption
spectrum of the system,\cite{Gavish6210637} has been demonstrated
experimentally.\cite{Deblock30120303,Zakka9923680307,Kouwenhoven9617660106,Billangeon10104109}
In recent years, the higher order current-correlations in a
nonequilibrium steady state are also explored both in
experimental and theoretical studies. \cite{Ubbelohde36122012,Zazunov990660107}

The above investigations were focused on the steady-state
transport regime. Whereas the theoretic development on transient
quantum transport dynamics,\cite{Jauho96, Schmidt08, Segal10,Tu11,Kennes12} there are of
considerable interests of the transient current fluctuation and
noise in the time domain. Recently, the transient current
fluctuation (correlation at equal time) of a two-probe transport
junction in response to the sharply turning off the bias voltage is
analyzed by Feng {\it et al.} \cite{Guo7707530208} The transient
evolution of finite-frequency current noise after abruptly switching
on the tunneling coupling in the resonant level model and the
Majorana resonant level model has also been studied by Joho {\it et
al.}\cite{Komnik8615530412} In this paper, we shall investigate the
transient current-current correlations of a biased quantum dot
system in the nonlinear transient transport regime.  Using the exact
master equation we developed
recently,\cite{Tu7823531108,Jin1208301310} a general formalism for
transient current-current correlations and transient noise spectra
are presented for an arbitrary spectral density of nanostructures.
This enables one to closely look at electron correlations during
transport in the time domain for the system not only in the steady
state but also when it is far away from the equilibrium. In
particular, it can unveil how the electron correlation changes in
the system when it evolves far away from the equilibrium to the
steady state, and the time-scale the system reaches the steady
state. These results should be useful for understanding the role
of quantum coherence and non-Markovian dynamics in quantum
transport. They are also essential for reconstructing quantum
states of electrons in nanostructures for further applications in
nanotechnology, such as the controlling of quantum information
processing and quantum metrology on quantum states, etc.

The rest of paper is organized as follows. In the next section,
transient current-current correlations in nanoelectronic systems are
formulated, and a general solution is presented using the master
equation formalism associated with the quantum Langevin equation. To
justify the correctness of our formalism, we examine the
steady-state current-current correlation of a single-level
nanostructure over the whole frequency range in
Sec.~\ref{noisespectrum}, in comparison with the results obtained
recently by Rothstein {\it et al.} \cite{Wohlman7907530709}. 
In Sec.~\ref{correlationsingle}, the
transient current-current correlations of the same system are
analyzed in details, and the energy structures of the noise spectra
are also explored. Conclusion is given in Sec.~\ref{conclusion}.

\section{Transient current-current correlations}
\label{DOTCCCUQLE}
To study the transient electronic transport and transient
current-current correlations in mesoscopic systems, we begin with 
Anderson impurity model. The Hamiltonian of the total system,
including the central dot, the leads and the coupling between them can be expressed as
\begin{align}
\label{Hamiltonian}
H = &\sum_{ij}\varepsilon_{ij}a_{i}^{\dag}a_{j}
+ \sum_{\alpha k}\epsilon_{\alpha k}c_{\alpha k}^{\dag}c_{\alpha
k} \notag\\
&+ \sum_{i\alpha k}[V_{i \alpha k}a^{\dag }_ic_{\alpha k} + {\rm
H.c.}],
\end{align}
and the electron-electron interaction is not considered.
Here $\varepsilon_{ij}$ and $\epsilon_{\alpha k}$ are the
corresponding energy levels of the dot and the lead $\alpha$, which
are experimentally tunable through the bias and gate voltages,
$V_{i\alpha k}$ is the tunneling
amplitude between the orbital state $i$ of the dot and the orbital
state $k$ of the lead $\alpha$, which can also be tuned by changing
tunneling barriers via external gate voltages, $a_{i}^{\dag}$ ($a_{i}$) and $c_{\alpha
k}^{\dag}$ ($c_{\alpha k}$) are creation (annihilation) operators of
electrons in the dot and the lead $\alpha$, respectively.

The current-current auto-correlation ($\alpha = \alpha'$)
and cross-correlation ($\alpha \neq \alpha'$) functions
are defined as follows,
\begin{align}
\label{currentcorrelationdef}
S_{\alpha \alpha'}(t+\tau, t)\equiv
\langle\delta I_{\alpha}(t+\tau)\delta I_{\alpha'}(t) \rangle,
\end{align}
where $\delta I_\alpha(t)\equiv I_\alpha(t)-\langle
I_{\alpha}(t)\rangle$ is the fluctuation of the current in the lead
$\alpha$ at time $t$. $I_\alpha$(t) is the current operator of
electrons flowing from the lead $\alpha$ into the central dot. It is
determined by
\begin{align}
\label{currentoperator}
I_{\alpha}(t) =& -e\frac{d}{dt}N_\alpha(t)=i\frac{e}{\hbar}[N_\alpha(t),H(t)]\notag\\
= &-i\frac{e}{\hbar}\sum_{ik}[V_{i\alpha k}a^{\dag}_i(t)c_{\alpha
k}(t)-V^{*}_{i\alpha k}c^{\dag}_{\alpha k}(t)a_i(t)],
\end{align}
where $e$ is the electron charge, $N_\alpha(t)=\sum_k c_{\alpha
k}^{\dag}(t)c_{\alpha k}(t)$ is the particle number operator of the
lead $\alpha$. The angle brackets in
Eq.~(\ref{currentcorrelationdef}) takes the mean value of the
operator over the whole system, which is defined as $\langle
O(t)\rangle = {\rm tr}[O(t)\rho_{\rm tot}(t_0)]$. Here $\rho_{\rm
tot}(t_0)$ is the initial state of the total system. Current-current
correlations measure the correlations between currents flowing in
different time. If we take Fourier transform of the current-current
correlation in Eq.~({\ref{currentcorrelationdef}}) with $\tau$,  an
asymmetric noise spectrum of the electronic transport at time $t$ is
obtained. Explicitly,
\begin{widetext}
\begin{align}
\label{currentcorrelation}
S_{\alpha\alpha'}(t+\tau,t) =
\frac{e^{2}}{\hbar^{2}}\sum_{ijkk'}\Big\{
&- V_{i\alpha k}V_{j\alpha'k'}[\langle a_{i}^{\dag}(t+\tau)c_{\alpha k}(t+\tau)a_{j}^{\dag}(t)c_{\alpha' k'}(t)\rangle - \langle a_{i}^{\dag}(t+\tau)c_{\alpha k}(t+\tau)\rangle\langle a_{j}^{\dag}(t)c_{\alpha' k'}(t)\rangle]\notag\\
&- V_{i\alpha k}^{*}V_{j\alpha'k'}^{*}[\langle c_{\alpha k}^{\dag}(t+\tau)a_{i}(t+\tau)c_{\alpha'k'}^{\dag}(t)a_{j}(t)\rangle - \langle c_{\alpha k}^{\dag}(t+\tau)a_{i}(t+\tau)\rangle\langle c_{\alpha'k'}^{\dag}(t)a_{j}(t)\rangle]\notag\\
&+ V_{i\alpha k}V_{j\alpha'k'}^{*}[\langle a_{i}^{\dag}(t+\tau)c_{\alpha k}(t+\tau)c_{\alpha'k'}^{\dag}(t)a_{j}(t)\rangle - \langle a_{i}^{\dag}(t+\tau)c_{\alpha k}(t+\tau)\rangle\langle c_{\alpha'k'}^{\dag}(t)a_{j}(t)\rangle]\notag\\
&+ V_{i\alpha k}^{*}V_{j\alpha'k'}[\langle c_{\alpha
k}^{\dag}(t+\tau)a_{i}(t+\tau)a_{j}^{\dag}(t)c_{\alpha'k'}(t)\rangle
- \langle c_{\alpha k}^{\dag}(t+\tau)a_{i}(t+\tau)\rangle\langle
a_{j}^{\dag}(t)c_{\alpha'k'}(t)\rangle]\Big\} .
\end{align}
\end{widetext}
We may simply denote $S=S^{(1)}+S^{(2)}+S^{(3)}+S^{(4)}$, corresponding to
the four terms respectively in Eq.~(\ref{currentcorrelation}).  We will explore the
contribution of each term to the transient noise spectra later.

Current-current correlations are in general complex, the physical
observables are related to its real or imaginary parts,
\begin{align}
S_{\alpha \alpha'}(t+\tau, t)& = S'_{\alpha \alpha'}(t+\tau, t) +
iS''_{\alpha \alpha'}(t+\tau, t),
\end{align}
where
\begin{subequations}
\begin{align}
S'_{\alpha \alpha'}(t+\tau, t)& = \frac{1}{2}\langle\{\delta
I_{\alpha}(t+\tau) , \delta I_{\alpha'}(t)\}\rangle \\
S''_{\alpha \alpha'}(t+\tau, t)& = \frac{1}{2i}\langle[\delta
I_{\alpha}(t+\tau) , \delta I_{\alpha'}(t)]\rangle
\end{align}
\end{subequations}
are directly proportional to the fluctuation function and the
response function, respectively, in the linear response
theory.\cite{Zwanzig01,Mazenko06}
On the other hand, we may introduce the total current-current
correlation defined by
\begin{align}
S(t+\tau,t)\equiv\langle\delta I(t+\tau)\delta I(t) \rangle,  \label{totcorrelation}
\end{align}
where the total current operator $I(t)$ is given by
\begin{align}
I(t)=aI_L(t)-bI_R(t),
\end{align}
and the coefficients satisfy the relation $a + b = 1$, associated to the
symmetry of the transport setup (e.g., junction capacitances). Then
Eq.~(\ref{totcorrelation}) can be written as
\begin{align}
S(t+\tau,t)=a^2S_{LL}(t+\tau,t)+b^2S_{RR}(
t+\tau,t)\notag\\-ab\big[S_{LR}(t+\tau,t)+S_{RL}(t+\tau,t)\big].
\end{align}
The usual total current-current correlation corresponds to $a=b=1/2$.
Taking different values of $a$ and $b$ can also give other
current-current correlations, such as the auto-correlation ($a=1, b=0$ or $a=0, b=1$), etc.

Now, we shall calculate exactly these correlation functions
in terms of the exact master equation we developed recently
for the investigation of transient quantum electron transports in nanostructures.
\cite{Tu7823531108,Jin1208301310} By
consider the central dot as an open system and the leads as its
environment, the exact master equation to describe the electron
dynamics in the dot system is given by
\begin{align}
\label{mastereq}
\frac{d\rho(t)}{dt} =& \frac{1}{i}[H'_S(t),\rho(t)] + \sum_{ij}\{\bm{\gamma}_{ij}(t)[2a_j\rho(t)a_i^{\dag}  \notag\\
& - a_i^{\dag}a_j\rho(t) - \rho(t)a_i^{\dag}a_j] + \bm{\widetilde{\gamma}}_{ij}(t)[a_i^{\dag}\rho(t)a_j \notag\\
& - a_j\rho(t)a_i^{\dag} + a_i^{\dag}a_j\rho(t) -
\rho(t)a_ja_i^{\dag}]\},
\end{align}
where $\rho(t)\equiv\rm{tr_R}[\rho_{tot}(t)]$ is the reduced density
matrix of the central dot. The initial state of the dot system is
assumed to be uncorrelated with the leads before the tunneling
couplings are turned on, namely,
$\rho_{tot}(t_0)=\rho(t_0)\otimes\rho_E(t_0)$. Here the dot can be
in any arbitrary initial state $\rho(t_0)$ but the leads are
initially at equilibrium:
$\rho_E(t_0)=\frac{1}{Z}e^{-\sum_{\alpha}\beta_{\alpha}(H_\alpha-\mu_{\alpha}N_{\alpha})}$,
and $\beta_{\alpha}=(1/k_BT_{\alpha})$ is the initial inverse
temperature of the lead $\alpha$. The first term in the master
equation describes the unitary evolution of electrons in the dot system, where
the renormalization effect after integrated out all the lead degrees of freedom
has been fully taken into account. The resulting
renormalized Hamiltonian is $H'_S(t) =
\sum_{ij}\varepsilon'_{ij}(t)a_i^{\dag}a_j$. The remaining terms
give the nonunitary dissipation and fluctuations induced by
backactions of electrons from the leads, and are described by the
dissipation and fluctuation coefficients $\bm{\gamma}(t)$ and
$\bm{\widetilde{\gamma}} (t)$, respectively. All those time-dependent
coefficients in Eq.~(\ref{mastereq}) are given explicitly by
\begin{subequations}
\label{coeffcient}
\begin{align}
\bm{\varepsilon}'(t) =& \frac{i}{2}[\dot{\bm{u}}(t,t_0)\bm{u}^{-1}(t,t_0)-\rm{H.c.}],\\
\bm{\gamma}(t) =&
-\frac{1}{2}[\dot{\bm{u}}(t,t_0)\bm{u}^{-1}(t,t_0)+\rm{H.c.}],\\
\bm{\widetilde{\gamma}}(t) =& \dot{\bm{v}}(t,t)-
[\dot{\bm{u}}(t,t_0)\bm{u}^{-1}(t,t_0)\bm{v}(t,t)+\rm{H.c.}],
\end{align}
\end{subequations}
where $\bm{u}(t,t_0)$ and $\bm{v}(t,t_0)$ are related to the
nonequilibrium Green's function of the dot system in the
Schwinger-Keldysh nonequilibrium
theory.\cite{Schwinger240761,Kadanoff62} These Green's functions
obey the following integro-differential Dyson equations,
\begin{subequations}
\label{greenfn}
\begin{align}
&\frac{d}{d\tau}\bm{u}(\tau,t_0) +
i\bm{\varepsilon}\bm{u}(\tau,t_0)+
\sum_{\alpha}\int_{t_0}^{\tau}d\tau'\bm{g}_{\alpha}(\tau,\tau')\bm{u}(\tau',t_0) = 0,\\
&\frac{d}{d\tau}\bm{v}(\tau,t) + i\bm{\varepsilon}\bm{v}(\tau,t) +
\sum_{\alpha}\int_{t_0}^{\tau}d\tau'\bm{g}_{\alpha}(\tau,\tau')\bm{v}(\tau',t) \notag \\
& ~~~~~~~~~~~~~~~~~~~~~~~~~~~~~~~=
\sum_{\alpha}\int_{t_0}^td\tau'\bm{\widetilde{g}}_{\alpha}(\tau,\tau')\bm{u}^{\dag}(\tau',t_0),
\end{align}
\end{subequations}
subject to the boundary conditions $\bm{u}(t_0,t_0)=1$ and
$\bm{v}(t_0,t)=0$ with $t_0\leq\tau\leq t$. Here, the self-energy
correlations from the lead to the central dot,
$\bm{g}_{\alpha}(\tau,\tau')$ and
$\bm{\widetilde{g}}_{\alpha}(\tau,\tau')$, are found to be
\begin{subequations}
\label{selfenergycorrelation}
\begin{align}
\bm{g}_{\alpha}(\tau,\tau') =&
\int\frac{d\omega'}{2\pi}\bm{J}_{\alpha}(\omega')e^{-i\omega'(\tau-\tau')},\\
\bm{\widetilde{g}}_{\alpha}(\tau,\tau') =&
\int\frac{d\omega'}{2\pi}\bm{J}_{\alpha}(\omega')f_{\alpha}(\omega')e^{-i\omega'(\tau-\tau')}.
\end{align}
\end{subequations}
In Eq.~(\ref{selfenergycorrelation}), the function $J_{\alpha ij}(\omega) = 2\pi\sum_kV_{i\alpha
k}V^*_{j\alpha k}\delta(\omega-\epsilon_{\alpha k})$ is an arbitrary spectral
density of the environment (the leads), and
$f_{\alpha}(\omega)=[e^{\beta_{\alpha}(\omega-\mu_{\alpha})}+1]^{-1}$
is the Fermi-Dirac distribution of lead $\alpha$ at initial time
$t_0$.

The above exact master equation can be connected to the exact quantum
Langevin equation for the dot operator. The later can be derived
formally from the Heisenberg equation of motion
\begin{align}
\label{quantumLangevineq}
\frac{d}{dt}a_i(t) =& -i\sum_j
\varepsilon_{ij} a_j(t) -\sum_{\alpha j}\int_{t_0}^t d\tau g_{\alpha
ij}(t,\tau)a_j(\tau)\notag\\
&-i\sum_{\alpha k} V_{i\alpha k} c_{\alpha
k}(t_0)e^{-i\epsilon_{\alpha k}(t-t_0)}.
\end{align}
In the above quantum Langevin equation, the first term is determined by
the evolution of the dot system itself, the second term is the
dissipation risen from the coupling to the leads, and the last term
is the fluctuation induced by the environment (the leads), and
$c_{\alpha k}(t_0)$ is the electron annihilation operator of the
lead $\alpha$ at initial time $t_0$. The time non-local correlation
function $g_{\alpha ij}(t,\tau)$ in Eq.~(\ref{quantumLangevineq}) is
also given by Eq.~(\ref{selfenergycorrelation}a), which
characterizes backactions between the dot system and the leads.
Because the quantum Langevin equation~(\ref{quantumLangevineq}) is
linear to $a_i$, its general solution can be written as
\begin{align}
\label{at}
a_i(t)=\sum_j u_{ij}(t,t_0)a_j(t_0)+F_i(t),
\end{align}
where $u_{ij}(t,t_0)$ is the same nonequilibrium Green's function of
Eq.~(\ref{greenfn}a) that determines the energy level renormalization
and dissipation in the dot system, as described by the master
equation. The noise operator $F_i(t)$ obeys the following equation,
\begin{align}
\frac{d}{dt}F_i(t)= -i &\sum_{j}\epsilon_{ij} F_j(t) -\sum_{\alpha
j}\int_{t_0}^t d\tau g_{\alpha ij}(t,\tau)
F_j(\tau)  \nonumber\\
 -i &\sum_{\alpha k}V_{i\alpha k} c_{\alpha k}(t_0)
 e^{-i\epsilon_{\alpha k}(t-t_0)}  \label{foe}
\end{align}
with the initial condition $F_i(t_0)=0$. Since the system and the leads are initially
decoupled to each other, and the leads are initially in equilibrium,
it can be shown that the solution of Eq.~(\ref{foe}) gives
\begin{align}
\langle F_j^{\dag}(t) &F_i(\tau)\rangle = v_{ij}(\tau,t)  \notag\\
&= \sum_{\alpha}\int_{t_0}^{\tau}dt_1\int_{t_0}^tdt_2
\big[\bm u(\tau,t_1)\widetilde{\bm g}_{\alpha}(t_1,t_2)
\bm u^{\dag}(t,t_2)\big]_{ij}, \label{v}
\end{align}
which is indeed the solution of Eq.~(\ref{greenfn}b). Thus the
connection of the solution of the quantum Langevin equation to
the fluctuation dynamics in the master equation is explicitly established.
Furthermore, the time-dependent operator $c_{\alpha k}(t)$ of the
lead $\alpha$ can also be obtained from its equation of motion:
\begin{align}
\label{ct}
c_{\alpha k}(t) =& c_{\alpha k}(t_0)e^{-i\epsilon_{\alpha k}(t-t_0)}\notag\\
&-i\sum_i \int_{t_0}^t d\tau V^*_{i\alpha k}
a_i(\tau)e^{-i\epsilon_{\alpha k}(t-\tau)}.
\end{align}
Using the solutions of Eq.~(\ref{at}) and (\ref{ct}), we can
calculate explicitly and exactly the current-current correlation
function (\ref{currentcorrelation}). The explicit expression is
still very complicated so we take the situation that the dot has no initial
occupation. Then, the four terms in  Eq.~(\ref{currentcorrelation}) is
given respectively by
\begin{widetext}
\begin{subequations}
\label{cccf}
\begin{align}
\label{1st} S_{\alpha \alpha'}^{(1)}(t+\tau,
t)=-\frac{e^2}{\hbar^2}{\rm
Tr}\Big\{&\big[\int_{t_0}^{t+\tau}ds\bm{g}_{\alpha}(t+\tau,s)\overline{\bm{v}}(s,t)-\int_{t_0}^{t}ds\bm{\widetilde{\bar{g}}}_{\alpha}(t+\tau,s)\bm{u}^{\dag}(t,s)\big]
\notag\\
\times&\big[\int_{t_0}^{t+\tau}ds'\bm{\widetilde{g}}_{\alpha'}(t,s')\bm{u}^{\dag}(t+\tau,s')-\int_{t_0}^{t}ds'\bm{g}_{\alpha'}(t,s')\bm{v}(s',t+\tau)\big]
\Big\},
\end{align}
\begin{align}
\label{2nd} S_{\alpha \alpha'}^{(2)}(t+\tau,
t)=-\frac{e^2}{\hbar^2}{\rm Tr}\Big\{&\big[\int_{t_0}^{t+\tau}ds\bm{v}(t,s)\bm{g}_{\alpha}(s,t+\tau)-\int_{t_0}^{t}ds\bm{u}(t,s)\bm{\widetilde{g}}_{\alpha}(s,t+\tau)\big]
\notag\\
\times&\big[\int_{t_0}^{t+\tau}ds'\bm{u}(t+\tau,s')\bm{\widetilde{\bar{g}}}_{\alpha'}(s',t)-\int_{t_0}^{t}ds'\overline{\bm{v}}(t+\tau,s')\bm{g}_{\alpha'}(s',t)\big]\Big\},
\end{align}
\begin{align}
\label{3rd} S_{\alpha \alpha'}^{(3)}(t+\tau,
t)=&+\frac{e^2}{\hbar^2}{\rm
Tr}\Big\{\big[\bm{\widetilde{\bar{g}}}_{\alpha}(t+\tau,t)\delta_{\alpha\alpha'}+\int_{t_0}^{t+\tau}ds\int_{t_0}^{t}ds'\bm{g}_{\alpha}(t+\tau,s)\overline{\bm{v}}(s,s')\bm{g}_{\alpha'}(s',t)
\notag\\
&-\int_{t_0}^{t+\tau}ds\int_{t_0}^{s}ds'\bm{g}_{\alpha}(t+\tau,s)\bm{u}(s,s')\bm{\widetilde{\bar{g}}}_{\alpha'}(s',t)-\int_{t_0}^{t}ds\int_{t_0}^{s}ds'\bm{\widetilde{\bar{g}}}_{\alpha}(t+\tau,s')\bm{u}^{\dag}(s,s')\bm{g}_{\alpha'}(s,t)\big]\bm{v}(t,t+\tau)\Big\},
\end{align}
\begin{align}
\label{4th} S_{\alpha \alpha'}^{(4)}(t+\tau,
t)=&+\frac{e^2}{\hbar^2}{\rm
Tr}\Big\{\overline{\bm{v}}(t+\tau,t)\big[\bm{\widetilde{g}}_{\alpha}(t,t+\tau)\delta_{\alpha\alpha'}+\int_{t_0}^{t+\tau}ds\int_{t_0}^{t}ds'\bm{g}_{\alpha'}(t,s')\bm{v}(s',s)\bm{g}_{\alpha}(s,t+\tau)
\notag\\
&-\int_{t_0}^{t+\tau}ds\int_{t_0}^{s}ds'\bm{\widetilde{g}}_{\alpha'}(t,s')\bm{u}^{\dag}(s,s')\bm{g}_{\alpha}(s,t+\tau)-\int_{t_0}^{t}ds\int_{t_0}^{s}ds'\bm{g}_{\alpha'}(t,s)\bm{u}(s,s')\bm{\widetilde{g}}_{\alpha}(s',t+\tau)\big]\Big\}.
\end{align}
\end{subequations}
\end{widetext}
Here, $\overline{v}_{ij}(\tau,t)=\langle
a_i(\tau)a_j^{\dag}(t)\rangle$, is related to the greater Green's
function in Keldysh's nonequilibrium approach. Its general solution
is given by
\begin{align}
\overline{\bm{v}}(\tau,t)=
\theta(\tau-t)\bm{u}(\tau,t)+\theta(t-\tau)\bm{u}^{\dag}(t,\tau)-\bm{v}(\tau,t).
\end{align}
The function $\bm{\widetilde{\bar{g}}}_{\alpha}(\tau,\tau')=\int\frac{d\omega'}
{2\pi}\bm{J}_{\alpha}(\omega')[1-f_{\alpha}(\omega')]e^{-i\omega'(\tau-\tau')}$
is a self-energy correlation of electron holes.
As one see, the transient current-current correlations have been expressed explicitly in terms of
our nonequilibrium Green's functions $\bm u(t,t_0)$ and $\bm v(\tau,t)$ that determine the
dissipation and fluctuation coefficients in the exact master equation (\ref{mastereq}).

\section{Noise spectra of a single-level nanostructure}
\label{noisespectrum}

 To justify the correctness of the above formalism, we first calculate
the noise spectra of a single level quantum dot coupled to two
leads over the whole frequency range, which has been recently investigated  in the
literatures.\cite{Engel9313660204,Wohlman7907530709}
The noise spectra can be obtained by taking the steady-state limit,
namely, setting
$t_{0}\rightarrow-\infty$ and let $t \rightarrow \infty$, and then
making a Fourier transformation to the total correlation (\ref{totcorrelation}).
This gives the  asymmetric noise spectrum as follows:
\begin{align}
S(\omega)\equiv\lim_{t\rightarrow\infty}\int_{-\infty}^{\infty}d\tau
e^{-i\omega \tau}\langle\delta I(t+\tau)\delta I(t)
\rangle,\label{noisedef1}
\end{align}
and it can be expressed by
\begin{align}
S(\omega)=a^2S_{LL}(\omega)+b^2S_{RR}(
\omega)-ab\big[S_{LR}(\omega)+S_{RL}(\omega)\big]\label{noisedef2}.
\end{align}
$S_{\alpha\alpha'}(\omega)$ in Eq.~({\ref{noisedef2}}) denotes the
auto-correlation noise ($\alpha=\alpha'$) and the cross-correlation one
($\alpha\neq\alpha'$):
\begin{align}
S_{\alpha
\alpha'}(\omega)\equiv\lim_{t\rightarrow\infty}\int_{-\infty}^{\infty}d\tau
e^{-i\omega \tau}\langle\delta I_{\alpha}(t+\tau)\delta
I_{\alpha'}(t) \rangle.\label{correlationnoisedef}
\end{align}
Since the entire system is in the steady-state limit, the current correlations
only depend on the time difference between measurements, it is clear
that $S_{\alpha\alpha'}(\omega)=S_{\alpha'\alpha}^{*}(\omega)$. This
relation makes the auto-correlation noise be real. We may define the
average cross-correlation noise $\bar{S}_{LR}(\omega)=
(S_{LR}(\omega)+S_{RL}(\omega))/2$, which is also real.
As one can see from Eq.~(\ref{noisedef2}) , the total noise spectrum is also real.

In the literature, one also assumes that the tunneling
couplings between the leads and the dot as well as the densities of
states of the leads are energy independent, i.e. the so-called
wide band limit (WBL). In our formalism, this corresponds to
$J_{\alpha}(\omega')=\Gamma_{\alpha}$, and the self-energy
correlations are reduced to:
\begin{align}
\label{selfenergycorrelation_WBL}
g_{\alpha}(\tau,\tau^{'})&=\Gamma_{\alpha}\delta(\tau-\tau^{'}),\notag\\
g^{\beta}_{\alpha}(\tau,\tau^{'})&=\Gamma_{\alpha}\int_{-\infty}^{\infty}
\frac{d\omega'}{2\pi}f_{\alpha}(\omega')e^{-i\omega'(\tau-\tau^{'})}.
\end{align}
The corresponding nonequilibrium Green's funciton are
\begin{align}
\label{uv_WBL}
u(\tau, t) = & e^{(i\epsilon+\frac{\Gamma}{2})(\tau-t)}, \notag\\
v(\tau, t) = & \int_{-\infty}^{\infty}\frac{d\omega'}{2\pi}\frac{\Gamma_Lf_L(\omega')+\Gamma_Rf_R(\omega')}{(\epsilon-\omega')^2+(\frac{\Gamma}{2})^2} \notag \\
& ~~~~~\times \big[e^{-i\omega'(\tau-t_0)}-e^{-(i\epsilon+\frac{\Gamma}{2})(\tau-t_0)}\big] \notag \\
& ~~~~~\times \big[e^{i\omega'(t-t_0)}-e^{-(-i\epsilon+\frac{\Gamma}{2})(t-t_0)}\big]
\end{align}
Substituting Eqs.~(\ref{selfenergycorrelation_WBL}) and (\ref{uv_WBL}) into
Eq.~(\ref{cccf}) and letting $t \rightarrow \infty$ (the steady-state limit), and then taking a
Fourier transform, we obtain the auto-correlation noise and the cross-correlation noise
for a single-level quantum dot system,
\begin{widetext}
\begin{align}
S_{\alpha\alpha}(\omega)=\frac{e^2}{\hbar^2}\int_{-\infty}^{\infty}
\frac{d\omega'}{2\pi}
\Big\{\Gamma_{\alpha}\Gamma_{\bar{\alpha}}\frac{f_{\alpha}(\omega')
\big[1-f_{\bar{\alpha}}(\omega'-\omega)\big]}{[\varepsilon-(\omega'-\omega)]^2
+(\Gamma/2)^2}+\Gamma_{\alpha}\Gamma_{\bar{\alpha}}
\frac{f_{\bar{\alpha}}(\omega')\big[1-f_{\alpha}(\omega'-\omega)\big]}
{(\varepsilon-\omega')^2+(\Gamma/2)^2}\notag\\
~~-\Gamma_{\alpha}^2\Gamma_{\bar{\alpha}}^2\frac{f_{\alpha}(\omega')
-f_{\bar{\alpha}}(\omega')}
{(\varepsilon-\omega')^2+(\Gamma/2)^2}\frac{f_{\alpha}(\omega'+\omega)
-f_{\bar{\alpha}}(\omega'+\omega)}{[\varepsilon-(\omega'+\omega)]^2+(\Gamma/2)^2}\notag\\
~~+\omega^{2}\Gamma_{\alpha}^{2}\frac{f_{\alpha}(\omega')}
{(\varepsilon-\omega')^2+(\Gamma/2)^2}\frac{1-f_{\alpha}(\omega'-\omega)}
{[\varepsilon-(\omega'-\omega)]^2+(\Gamma/2)^2}
\Big\}, \notag\\
\notag \\
\bar{S}_{LR}(\omega)=\frac{e^2}{\hbar^2}\int_{-\infty}^{\infty}
\frac{d\omega'}{2\pi}
{\rm Re} \Big\{-\frac{\Gamma_{L}\Gamma_{R}}
{-i(\varepsilon-\omega')+\Gamma/2}\frac{f_{R}(\omega')\big[1-f_{L}(\omega'-\omega)]+f_{L}(\omega')\big[1-f_{R}(\omega'-\omega)\big]}
{i[\varepsilon-(\omega'-\omega)]+\Gamma/2}\notag\\
+\Gamma_{L}^2\Gamma_{R}^2\frac{f_{L}(\omega')-f_{R}(\omega')}
{(\varepsilon-\omega')^2+(\Gamma/2)^2}\frac{f_{L}(\omega'+\omega)
-f_{R}(\omega'+\omega)}{[\varepsilon-(\omega'+\omega)]^2+(\Gamma/2)^2}
\Big\}.
\end{align}
\end{widetext}
The above two equations provide
the exact noise spectra at finite temperature and finite bias over the entire
frequency range in the WBL. The noise spectrum is proportional to the
emission-absorption spectrum of the system, so $S(\omega)$ can be
viewed as the probability of a quantum energy $\hbar\omega$ being
transferred from the system to a measurement apparatus.

\begin{figure}
\includegraphics[width =8.5 cm]{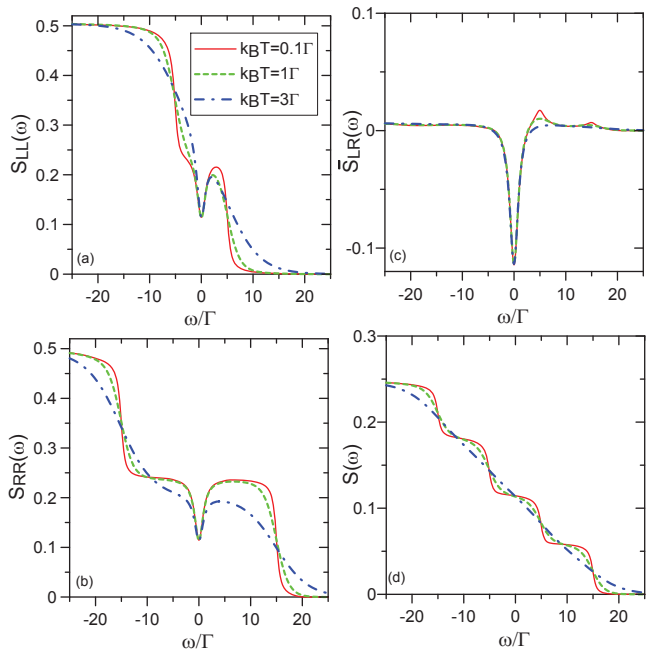}
\caption{Steady-state noise spectra (in unit of $e^2 \Gamma / \hbar^2$) of the
current transport through a single-level quantum dot as a function of
the detected frequency $\omega$, where
$\varepsilon=5\Gamma$ with $\Gamma_L=\Gamma_R=0.5\Gamma$ at the
bias $eV=20\Gamma$. Difference curves correspond to different initial
temperatures of the leads as shows in the figures. } \label{NoiseTdep}
\end{figure}
In Fig.~{\ref{NoiseTdep}}, we plot the noise spectra at several different
initial temperatures under a fixed bias. In all the plots, we
set $\mu_L+\mu_R=0$ so that $\mu_L=-\mu_R=eV/2$.
The noise spectra clearly shows an asymmetric structure.
For auto-correlation noises (see Fig.~\ref{NoiseTdep}(a) and (b)),
there are step structures (except for frequency near zero) as a
function of frequency, with the step edges located roughly at the
resonant tunneling frequencies
$\omega_\alpha=|\mu_\alpha-\varepsilon|= \mp 5 \Gamma$ ($\mp
15\Gamma$) for the left (right) lead at temperature
$k_BT=0.1\Gamma$, as evidences of  sequential tunnelings. The step
height saturates at
$S_{\alpha\alpha}(\omega)=e^2\Gamma_{\alpha}/\hbar^2$. The plateaus
in the step structure come from the WBL approach (the result goes
beyond the WBL will be presented in the next section, combining with
the analysis of the transient noise spectra). They correspond to
those events in which electron transport processes are independent
to each other (uncorrelated).  As a result, the tunneling rate and
the level population can be extracted from the step heights and the
ratio of step heights, respectively.\cite{Engel9313660204}  When the
initial temperature increases, the thermal broadening effect near
the fermi surface of the leads smears the resonance effects, which
causes the step structure to vanish.  The Lorentzian dips in the two
auto-correlation noises around $\omega=0$ are associated with
negative correlations between the currents. Since here the
electron-electron interaction inside the dot is not considered, the
dips are purely due to the Pauli exclusion
principle.\cite{Chen43453491} The average cross-correlation noise
(see Fig.~\ref{NoiseTdep}(c)) has a much smaller scale, in
comparison with the auto-correlation noise. This is reasonable
because the cross-correlation is established between the two leads
through the central dot. In Fig.~\ref{NoiseTdep}(c), there are two
peaks at the resonant tunneling frequencies $\omega_{L, R}=|\mu_{L,
R}-\varepsilon| = 5\Gamma$  and $15 \Gamma$ when the initial
temperature is low. Again, when the temperature gets higher, the
effect is smeared. The peaks are associated with a positive
correlation of the currents between the initially uncorrelated left
and right leads. There is also a Lorentzian dip in the
cross-correlation noise at zero frequency with the same reason as
the auto-correlation noises, so that the scale of the dips are the
same. For a symmetric transport setup ($a=b=1/2$), the total net
current noise (see Fig.~\ref{NoiseTdep}(d)) has four steps,
corresponding to the two resonances in each frequency side, but
there is no Lorentzian dip around zero frequency. This is because
the dips in the auto-correlation noises and in the cross-correlation
noises are canceled each other in the symmetric setup.  These
results coincide with the results obtained recently by Rothstein
{\it et al.} \cite{Wohlman7907530709} using the scattering matrix
approach.

\begin{figure}
\includegraphics[width =8.5 cm]{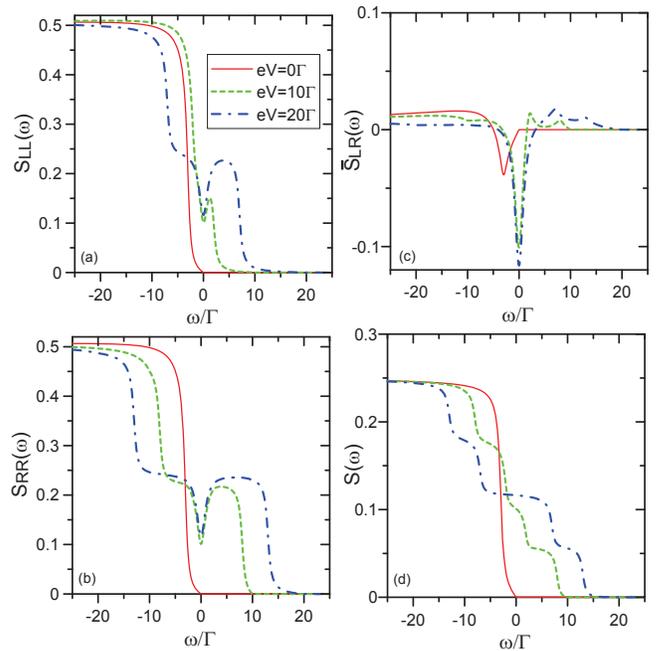}
\caption{Steady-state noise spectra (in unit of $e^2 \Gamma / \hbar^2$) of the
current transport through a single-level quantum dot as function of
detected frequency $\omega$ at zero
temperature with different bias voltages, where
$\varepsilon=3\Gamma$, with $\Gamma_L=\Gamma_R=0.5\Gamma$. } \label{NoiseVdep}
\end{figure}
  Fig.~{\ref{NoiseVdep}} concerns noise spectra under
different biases applying to the leads. To keep away with the
thermal noise, we set the temperature of the two leads to be zero,
that is, the whole system is in the shot noise regime. We can see
that the noise spectra would exactly be zero when the detected
frequency is larger than the bias voltage. This means that no
tunneling process can happen in this regime. For auto-correlation
noises (See Fig.~\ref{NoiseVdep}(a) and Fig.~\ref{NoiseVdep}(b)), we
still have step structure with the step edges locating at the
resonant tunneling frequencies
$\omega_{\alpha}=|\mu_{\alpha}-\varepsilon|$ for both the positive
and negative frequency axes except for zero bias ($eV=0\Gamma$). At
zero bias, the noise spectrum only has one step edged at negative
resonant tunneling frequency. For the cross-correlation noise (see
Fig.~\ref{NoiseVdep}(c)), the dip is shifted to the resonant
frequency without a bias ($eV=0\Gamma$), and the scale of the dip is
smaller than that in biased cases. When we increase the bias, saying
$eV=10\Gamma$, in additional to the two peaks at the resonant
frequency, we find a small dip around the bias voltage in the
negative frequency axis. We speculate that this is an evidence of
cotunneling processes. The cotunneling dip is less obvious because
the cotunneling events should have a much smaller probability, in
comparison with the sequential tunnelings.

\section{Transient Current-Current Correlations of the Single
Level Nanostructure}
\label{correlationsingle}

Now we shall study the transient current-current
correlations of the system that we discussed in the last section.
For the sake of generality, we extend the spectral density to an
energy-dependent one. We assume that the electronic structure of the
leads has a Lorentzian line shape
\cite{Mac06085324,Jin08234703,Jin1208301310,Tu7823531108}.
\begin{align}
J_{\alpha}(\omega)=\frac{\Gamma_{\alpha}W_{\alpha}^2}{(\omega-\mu_{\alpha})^2+W_{\alpha}^2},
\end{align}
where $W_{\alpha}$ is the band width and $\Gamma_{\alpha}$ is the
coupling strength of lead $\alpha$. The current-current correlation
describes how the correlation persists until it is averaged out through the
coupling with the surroundings. Thus, we fix the observing time $t$,
and see how the correlation varies via the time difference $\tau$ of measurements.
We take the initial time $t_0=0$.
\begin{figure}
\includegraphics[width =5.5 cm]{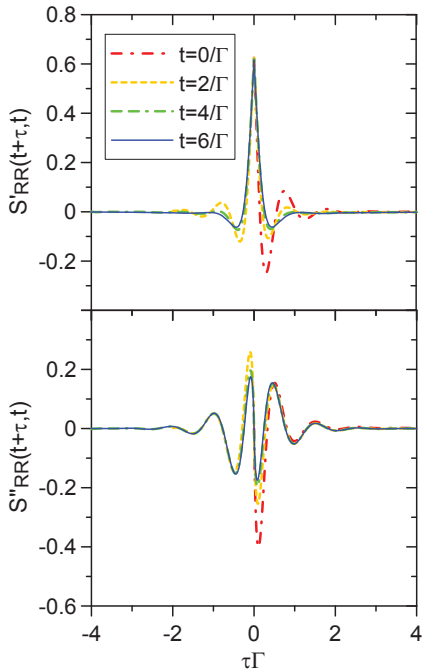}
\caption{Auto-correlation function $S_{RR}$ 
in terms of their real and imaginary parts 
(in units of $e^2\Gamma^2/\hbar^2$) in a single-level nanostructure for different $t$ as
a function of $\tau$. Where $\varepsilon=\Gamma$, with
$\Gamma_L=\Gamma_R=0.5\Gamma$, $W_L=W_R=5\Gamma$, $eV=10\Gamma$, at
$k_{B}T=0.5\Gamma$ for both two leads.}
\label{correlaitons_difft}
\end{figure}
In Fig.~\ref{correlaitons_difft}, we plot the auto-correlation
function for several different $t$. This allows one to monitor the
transient processes until the system reaches its steady state at
which these correlations come to only depend on the time difference
$\tau$ between the measurements. As one can see both the real part
and imaginary part of correlation functions approach the
steady-state values at $t \simeq 5/\Gamma$. The real part of the
auto-correlation has a maximal value at $\tau=0$ (namely when it is
measured in the same time), this gives the current-fluctuation, and
we find that this current-fluctuation is independent of the
observing time $t$ (less transient). When the time difference $\tau$
gets larger, the auto-correlation decays rather faster, and it
reaches to zero after $\tau > 2/\Gamma$, namely the correlation
vanishes.
With the observing time goes on, the real part of auto-correlation
becomes more and more symmetric, and the imaginary part gets
more antisymmetric, and eventually they becomes fully symmetric
and antisymmetric function of $\tau$, respectively, in the steady-state limit, as one expected.
We also find that the cross-correlation
is rather small (about of one order of magnitude smaller in comparison with the
auto-correlation) so that it is not presented in Fig.~\ref{correlaitons_difft}.

\begin{figure}
\includegraphics[width =10 cm]{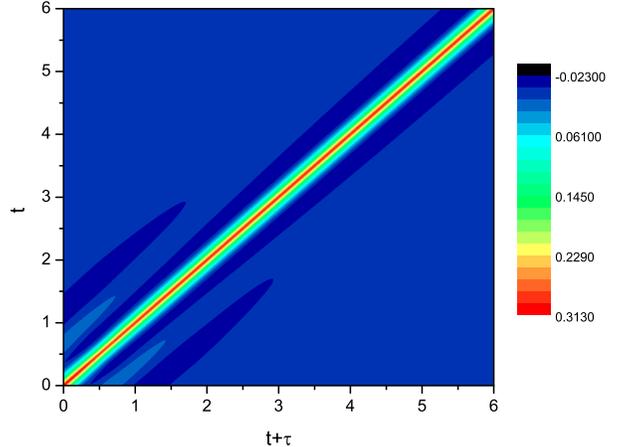}
\caption{The contour plot of the real part of the total current-current correlation,
$S'(t+\tau,t)$ (in units of $e^2\Gamma^2/\hbar^2$), in the single-level nanostructure
in the two-time plane (scaled by $\Gamma$).
Here the parameter $\varepsilon=\Gamma$, with $\Gamma_L=\Gamma_R=0.5\Gamma$,
$W_L=W_R=5\Gamma$, $eV=10\Gamma$, at $k_{B}T=0.5\Gamma$ for both two
leads.} \label{contour_plot}
\end{figure}

To have a more general picture how the system reaches the steady
state, we present a contour plot of the real part of the
total-correlation in the 2-D time domain in Fig~\ref{contour_plot}.
As one see it is symmetric in the diagonal line ($\tau=0$), as a
consequence of the identity: $S_{\alpha \alpha'}(t+\tau,
t)=S^{*}_{\alpha' \alpha}(t, t+\tau)$.  The contour-plot clearly
shows an oscillating profile of the correlation in the region
$t<3/\Gamma$. The oscillation quickly decays for the time $3/\Gamma
< t < 5/\Gamma$. The correlation reaches a steady-state value after
$t  \simeq 5/\Gamma$. The imaginary part has much the same behavior,
except that it has an antisymmetric profile in terms of $t$ and
$t+\tau$. This gives the whole picture of the transient
current-current correlation.

\begin{figure}
\includegraphics[width =5.5 cm]{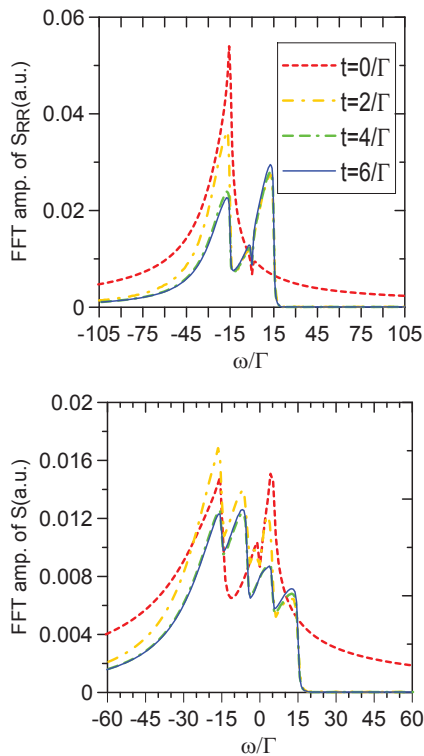}
\caption{The FFT Amplitude of the auto-correlation $S_{RR}$ and
the total correlation $S$ in the single-level nanostructure
as a function of $\omega$ (in units of $\Gamma$). Where
$\varepsilon=5\Gamma$, $\Gamma_L=\Gamma_R=0.5\Gamma$,
$W_L=W_R=15\Gamma$, $eV=20\Gamma$, $k_{B}T=0.1\Gamma$ for both two
leads.  } \label{Amp}
\end{figure}

To see the energy structure in electron transports through the transient current-current correlations,
we use the fast Fourier transform (FFT) to convert the correlation functions from the time
domain ($\tau$) into the frequency domain for different observing time $t$. The result gives
the standard definition of the transient noise spectra. In Fig.~\ref{Amp},
we plot the FFT amplitude of auto-correlation
$S_{RR}(t+\tau, t)$ and total-correlation $S(t+\tau, t)$.
From Fig.~\ref{Amp}, one can analyze the electron transport
properties through the noise spectra not only just in the steady state,
but also in the entire transient regime.
To make the energy structure manifest in the transient noise spectra,
we let the initial temperature approaches to zero ($k_BT=0.1\Gamma$).
The right auto-correlation shows only one single peak at $\omega_-=-\omega_R=-|\mu_R-
\varepsilon|$ in the beginning. This is because the dot is initially empty so that
electrons tunneling from the Fermi surface of the right lead to the dot have
a maximum probability. This peak corresponds to the energy absorption
of the electron tunnelings. On the other hand, we also observed that the
tunneling process for $\omega>eV$, which is forbidden in the steady
state near zero temperature, can happen in the transient regime. As
the time $t$ varies, the second peak shows up. This comes from backward electron
tunnelings (i.e. emission processes) from dot to the right lead, with the peak edge locating at the
resonance frequency $\omega_+=\omega_R=|\mu_R-\varepsilon|$.
Note that with a finite bandwidth spectral density,
the spectrum decays when the frequency passes over the resonant
frequencies, which is different from the WBL where spectrum is flat.
The noise spectrum still has a dip at zero frequency in both the transient and steady-state
regimes. The FFT amplitude of the total-correlation has
the same properties as the right auto-correlation, with two more peaks
coming from the left auto-correlation functions as effects of the emission
and absorption processes between the left lead and the central dot.
By calculating the individual contribution of the four
terms in the auto-correlation expression (Eq.~(\ref{1st})-(\ref{4th})), we
find that $S^{(3)}$ and $S^{(4)}$ dominate the tunnelings from the
dot to the leads and via versus. The contributions from $S^{(1)}$ and $S^{(2)}$
are much smaller because they describe the correlations of electron tunneling
in the same direction, and mostly contribute to the noise around zero frequency
due to the Pauli exclusion principle.

\begin{figure}
\includegraphics[width =5.5 cm]{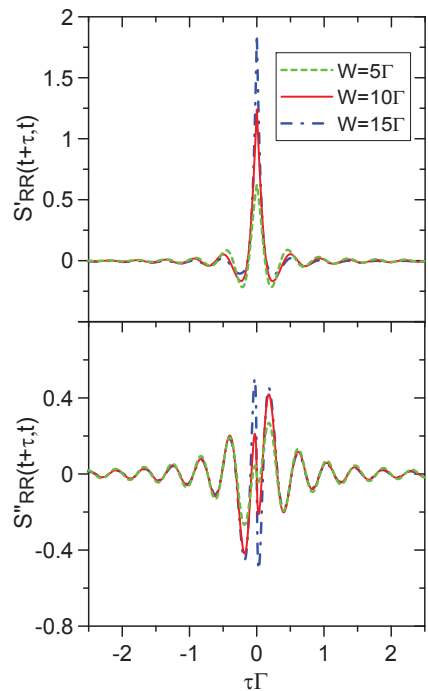}
\caption{Auto-correlation function (in units of
$e^2\Gamma^2/\hbar^2$) of the single-level nanostructure for
different band width $W=W_L=W_R$, at observing time $t=6/\Gamma$.
Where $\varepsilon=5\Gamma$, with $\Gamma_L=\Gamma_R=0.5\Gamma$,
$eV=20\Gamma$, and $k_BT=0.1\Gamma$ for both two leads.}
\label{auto_correlation_diffW}
\end{figure}

\begin{figure}
\includegraphics[width =5.5 cm]{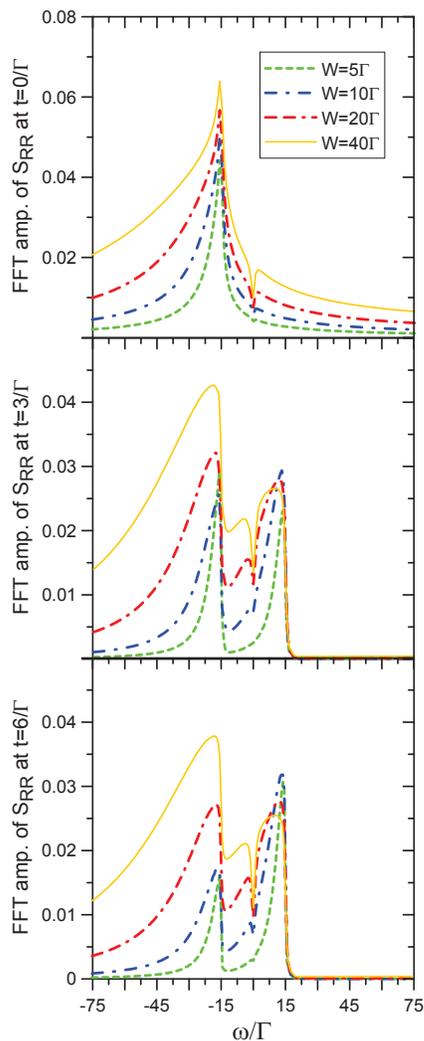}
\caption{The FFT Amplitude of $S_{RR}$ (in arbitrary unit) in the
single-level nanostructure for different band width
$W=W_L=W_R$, at three different observing time t. Where
$\varepsilon= 5\Gamma$, $\Gamma_L=\Gamma_R=0.5\Gamma$,
$eV=20\Gamma$, and $k_{B}T=0.1\Gamma$ for both two leads.  }
\label{Amp_diffW}
\end{figure}

In the end, we investigate how the
current-current correlation changes for different band widthes of
the spectral density. We first examine such changes in the time domain (see
Fig.~\ref{auto_correlation_diffW}). For the real part of the
auto-correlation, the notable difference is manifested in the current
fluctuation ($\tau = 0$). The current-fluctuation is increased with the increase of
the band width, and goes to infinity at WBL. For the imaginary part, the
amplitude of oscillation increases with the increase of the band width
only for $ |\tau| < 0.5/\Gamma$. For the transient noise spectra (see Fig.~\ref{Amp_diffW}),
we see that the tails of the correlation decay more and more slowly with
increasing the band width at $t=0$.  As the time $t$ varying, besides
the occurrence of another peak mentioned before, we notice that the dip
at zero frequency gradually vanishes as the band width is decreased.
When the system reaches to the steady state, the band-width dependence
of the noise spectra over the whole frequency range is clearly shown up.
These results could be examined easily in experiments.

\section{Conclusion}\label{conclusion}
In this paper, we present an exact formulation of transient
current-current correlations and transient noise spectra, and also
carry out the exact solution to Anderson impurity model, 
but the electron-electron interaction has not been included.
By inspecting the transient current-current correlations, we obtain
the information of electron transport processes before the system
reaching the steady state. We find that current-current
correlations have stronger signals in transient regime.
It provides interesting results on nonequilibrium
two-time correlation functions in nanostructures that could be
measured in experiments. Indeed, two-time correlation functions 
have been proposed and experimentally measured in optical
measurements,\cite{Malik98,Livet00} which provided further
information on microstructure of materials and the interplay between
the material's structure and transport properties. An more recent
application is in photosynthetic systems, in which one measured the
two-dimensional electronic spectroscopy through the Fourier
transform of  two-time correlation functions, to understand the high
efficiency of light-harvesting associated with possible
non-Markovian coherence energy transfer.\cite{Fleming07,Collini10}
We expect that similar measurements on two-time current
correlation functions could be done in nanostructured
systems, which should provide more useful information for the
controlling of nanosystems, in particular, for applications on quantum
information processing and quantum metrology of quantum states.

\section*{ACKNOWLEDGMENT}
We thank Amnon Aharony and Ora Entin-Wohlman for fruitful
discussions and useful comments on the manuscript. This work is supported
by the National Science Council of Republic of China under Contract
No. NSC 102-2112-M-006-016-MY3.


\end{document}